\documentclass[12pt,a4paper]{article}

\usepackage[margin=1in]{geometry}
\usepackage{setspace}
\usepackage{graphicx}
\usepackage{amsmath, amssymb}
\usepackage{hyperref}
\usepackage{natbib}
\usepackage{enumitem}
\usepackage{mathtools}
\usepackage{algorithm}
\usepackage{algpseudocode}
\usepackage{booktabs}
\usepackage{graphicx}
\usepackage{titling} 
\usepackage{graphicx}  
\usepackage{authblk}   
\usepackage{caption}   
\usepackage{siunitx}
\title{\textbf{Toward Black–Scholes for Prediction Markets: A Unified Kernel and Market-Maker’s Handbook}}
\author{
Shaw Dalen\footnote{shawdalen@daedalus-research.com}\\
\raisebox{-0.3ex}{\includegraphics[height=1.1em]{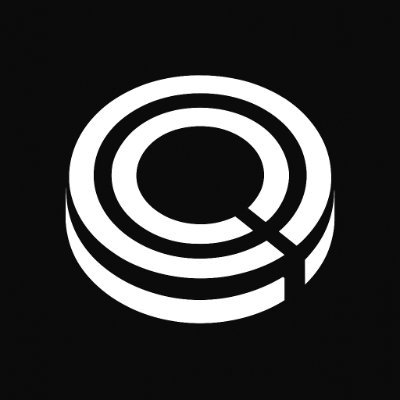}}\;\;Daedalus Research Team\footnote{\texttt{https://x.com/DaedalusRsch}}
}

\begin{document}
\date{} 
\maketitle

\begin{abstract}
Prediction markets---exemplified by Polymarket and similar venues---aggregate dispersed information into tradable probabilities, yet they still lack the unifying stochastic kernel that options gained from Black--Scholes. As these markets scale (institutional participants, exchange integrations, and rising volumes around elections and macro prints), makers face belief–volatility, jump, and cross–event risks without standardized tools to quote or hedge them. We propose such a foundation: a \emph{logit jump--diffusion with risk–neutral (RN) drift} that treats the traded probability $p_t$ as a $\mathbb{Q}$–martingale and exposes belief volatility, jump intensity, and dependence as quotable risk factors. On top, we build a calibration pipeline that filters microstructure noise, separates diffusion from jumps via EM, enforces the RN drift, and yields a stable belief–volatility surface. We then define a coherent derivative layer—variance, correlation, corridor, and first–passage instruments—analogous to volatility and correlation products in option markets. In controlled experiments (synthetic RN-consistent paths and real event data), the RN–JD model achieves lower short–horizon belief-variance forecast error than diffusion-only and $p$-space baselines, validating both its causal calibration and economic interpretability. Conceptually, the RN–JD kernel supplies the ``implied volatility'' analogue for prediction markets: a tractable, tradable language for quoting, hedging, and transferring \emph{belief risk} across venues such as Polymarket.

\end{abstract}

\section{Introduction}
\paragraph{What is prediction market and why we need it?}
Prediction markets like Polymarket trade contracts that pay \$1 if an event occurs and \$0 otherwise. Under standard no-arbitrage reasoning, observed prices are interpretable as \emph{risk-neutral} probabilities of the event. Empirically, these prices often track average beliefs under mild conditions, although biases exist and interpretation requires care \citep{WolfersZitzewitz2004,WolfersZitzewitz2006,Manski2004}. These markets have been used to forecast outcomes in politics, economics, science, and beyond \citep{Arrow2008,Snowberg2012}. 
Beyond entertainment or wagering, liquid, global event markets instantiate a Hayekian mechanism for aggregating dispersed knowledge \citep{Hayek1945}. Prices synthesize private signals into public probabilities, creating incentives for informed actors to reveal information when they expect to profit. Empirically and in practice, prediction markets can improve forecast accuracy and organizational decision-making by rewarding correct beliefs and penalizing noise \citep{CowgillZitzewitz2015}. The appeal has long been recognized by policymakers: the U.S. Defense Advanced Research Projects Agency (DARPA) explored a \emph{Policy Analysis Market} (FutureMAP) for geopolitical risks before cancelling it in 2003 amid political criticism \citep{WaPo2003,Hanson2007PAM}. At the same time, regulatory choices shape what can be listed: for example, the 1958 \emph{Onion Futures Act} still bans U.S. onion futures, illustrating how legal constraints can limit the scope of information aggregation via markets \citep{USC7_13_1}. Finally, recent debates over polling errors (e.g., the 2016 U.S. election) underscore the value of complementary, market-based signals when traditional information channels are noisy or biased \citep{AAPOR2016}. Together, these strands motivate a standardized kernel and derivative layer so that “belief risk’’ can be quoted, hedged, and transferred at scale.

\paragraph{Plain-language view of event contracts.}
An \emph{event contract} is a yes/no claim with a fixed payoff: it pays \$1 if the specified outcome occurs by a given date and \$0 otherwise. The traded price $p_t\!\in\!(0,1)$ is naturally read as the market’s risk-neutral probability for that outcome. Examples include ``Will the unemployment rate exceed 5\% in Q4?'' or ``Will upgrade $U$ activate by block height $H$?''. In regulated settings, the U.S. Commodity Futures Trading Commission (CFTC) describes such claims as derivative contracts tied to the occurrence of an event, typically with binary payoff structure; exchanges listing them must satisfy the same substantive requirements as other derivatives \citep{FedRegisterEventContracts2024,CFTCEventContracts}. Crypto-native venues list economically similar contracts under on-chain execution.

Execution is fragmented. Platforms rely on the \emph{Logarithmic Market Scoring Rule (LMSR)} \citep{Hanson2003}, on \emph{constant-product automated market makers (CP-AMMs)}—a special case of \emph{constant-function market makers (CFMMs)}—popularized by Uniswap \citep{UniswapV2,AngerisCFMM}, or on traditional order books. Each mechanism sets prices, but none provides a shared stochastic \emph{kernel} that explains how event probabilities should evolve over time, across information shocks, or jointly across related events. By contrast, once the Black--Scholes (BS) model appeared in options, markets standardized around \emph{implied volatility}, which enabled quoting, hedging, and a deep derivative layer \citep{BlackScholes1973}. Prediction markets lack an analogous foundation.

\paragraph{Motivation and background.}
Market making in event contracts is hard. In microstructure, informed trading induces \emph{adverse selection}: when a counterparty trades only when you are wrong, you lose on average \citep{GlostenMilgrom1985,Kyle1985}. In binary event markets, inventory risk is concentrated near resolution and cannot be hedged by the underlying until settlement. For cost-function market makers such as LMSR, providing more liquidity increases worst-case loss; without external subsidies, they are expected to run at a deficit proportional to the liquidity they offer \citep{OthmanSandholm2010}. CP-AMMs/CFMMs face related profitability constraints for liquidity providers \citep{AngerisUniswap2019,Bitterli2023}. In short, today’s mechanisms expose makers to toxic flow and gap risk, but do not offer standardized tools to transfer \emph{belief risk} (the risk that the market-implied probability moves) across time or across related events.

\paragraph{Why a derivative layer for event contracts?}
Derivatives exist to let participants isolate and trade specific risks. For events, the primary risks are movements in the \emph{belief level} and its \emph{volatility} (how fast log-odds move), plus jump risk from news and cross-event co-movement. A derivative layer would (i) let market makers hedge adverse selection by offloading belief volatility and jump exposure; (ii) allow calendar hedges (between maturities or checkpoints before resolution); (iii) enable cross-event hedges that neutralize correlation and co-jumps; and (iv) concentrate liquidity around a small set of quoted risk factors, as implied volatility did for options. In mature option markets, variance/volatility swaps, correlation swaps, and related instruments serve exactly these functions for price volatility \citep{Demeterfi1999,CarrLee2009,Broadie2008}. Prediction markets need the analogous instruments for belief dynamics.

\paragraph{Context: the rise of Polymarket.}
Crypto-native platforms have brought event trading to a wider audience and catalyzed institutional interest. In 2025, Intercontinental Exchange (ICE), the parent of the New York Stock Exchange, announced a strategic investment of up to \$2\,billion for roughly a 20\% stake in Polymarket, with plans to distribute event-driven data through ICE’s channels \citep{AxiosICE2025}. Separate disclosures report that Polymarket raised over \$200\,million across 2024–2025 prior to the ICE deal \citep{TheBlock205M2025}. Together with growth spurts around major elections and macro events, these developments signal mainstream acceptance of event markets and amplify the need for a common pricing kernel and a standardized derivative layer.

\paragraph{This paper.}
We propose a minimal, actionable kernel: a \emph{logit jump–diffusion with multi-event correlation}. Let $p_t\!\in(0,1)$ be the risk-neutral event probability and $x_t=\log(p_t/(1-p_t))$ its log-odds. We model $x_t$ as a correlated jump–diffusion. Enforcing the martingale property of $p_t$ under the risk-neutral measure pins down the drift of $x_t$; what remains—the belief-volatility $\sigma_b$, jump intensity and moments, correlation across events, and co-jump structure—are the tradable risk factors. On this kernel we define a coherent menu of \emph{event-linked derivatives} (belief variance/volatility swaps, correlation swaps, corridor variance, threshold/path notes, and conditional baskets). Where possible we give closed-form or short-maturity approximations; otherwise we use \emph{partial integro–differential equations (PIDE)} or \emph{Monte Carlo (MC)}. We derive Greeks with respect to $x$ and to the kernel’s parameters, and we outline practical hedges (calendar, cross-event, and inventory-aware rules near the $0/1$ boundaries). Finally, we describe a data-driven calibration pipeline that maps mid/bid–ask/trade data to a smoothed belief-volatility surface with co-jump detection.

\paragraph{Why now.}
Standardization matters more than perfect realism. Black--Scholes was not ``true,'' but it coordinated quoting and hedging around a small number of state variables; that standard enabled scale. A shared belief–variance surface can play the same role for event markets precisely as adoption accelerates. In 2025, Intercontinental Exchange (ICE), owner of the NYSE, announced \emph{up to} \$2\,billion of strategic investment in Polymarket (at roughly \$8\,billion pre-money) and will distribute its event-driven data to institutions, a signal of mainstream integration \citep{ICE2025PR,FT2025Polymarket,Reuters2025Polymarket}. On the usage side, monthly volumes have broken through billion-dollar thresholds: reports based on Dune/DeFiLlama data indicate Polymarket posted about \$1.4\,billion in September 2025 while Kalshi exceeded \$1–3\,billion depending on the week and product mix, with sports driving much of the surge \citep{Defiant2025PM,Decrypt2025Kalshi,LSR2025Kalshi}. At the same time, U.S. oversight is becoming more explicit: the CFTC proposed rulemaking on event contracts in 2024 (clarifying categories that may be contrary to the public interest), and litigation around political markets underscores an evolving but increasingly articulate framework \citep{CFTC2024NPRM,FR2024EventContracts,KalshiCADC2024}. This combination of institutional capital, record volumes, and regulatory clarity strengthens the case for a common pricing kernel and a standardized derivative layer to concentrate liquidity, reduce maker losses via hedging, and support an institutional market for belief risk.

\section{Related Work}

\subsection{Prediction markets and mechanisms.}
The \emph{Logarithmic Market Scoring Rule (LMSR)} introduced a bounded-loss, always-on \emph{automated market maker (AMM)} for event contracts and combinatorial claims, giving crisp axiomatic guarantees and a tractable \emph{cost function} representation \citep{Hanson2003,ChenPennock2012}. Subsequent designs generalized AMMs to convex cost-function markets and to liquidity-adaptive variants that mitigate worst-case losses and over-movement in thin markets \citep{Abernethy2013,OthmanSandholm2010}. On crypto rails, \emph{constant-function market makers (CFMMs)}---including the \emph{constant-product AMM (CP-AMM)} popularized by Uniswap---standardized execution and on-chain pricing primitives \citep{AngerisChitra2020,UniswapV2}. In contrast, \emph{central limit order books (CLOBs)} provide deep execution microstructure but no common probabilistic dynamics for event prices \citep{Gould2013}. Our focus is complementary: we seek a shared \emph{stochastic kernel} for risk-neutral event probabilities across time, shocks, and related events, independent of the execution venue.

\subsection{Option pricing, implied surfaces, and coordination role.}
The Black–Scholes–Merton paradigm established a common language for quoting and hedging (implied volatility, Greeks) \citep{BlackScholes1973}. It catalyzed successive layers: jumps \citep{Merton1976}, stochastic volatility \citep{Heston1993}, and implied-consistent \emph{local volatility} surfaces \citep{Dupire1994}, leading to the modern practice of surface construction and management \citep{Gatheral2006}. This standardization coordinated liquidity and risk transfer. Our work aims for the analogous role in prediction markets: replace price diffusions by \emph{logit} dynamics for probabilities on $(0,1)$, preserving tractability while exposing belief-level, belief-volatility, and jump/correlation factors as quotable objects.

\subsection{Information-based processes.}
Information-based asset pricing treats prices as conditional expectations under a filtration generated by noisy “information processes,” often Brownian-bridge–driven, offering a structural view of how news reveals payoffs over time \citep{BrodyHughstonMacrina2007}. Separately, boundary-constrained diffusions on $[0,1]$ (e.g., Wright–Fisher/Jacobi families) provide mathematically consistent dynamics for probabilities or bounded state variables \citep{JenkinsSpano2017,Ackerer2016}. We adopt a simple alternative: a \emph{logit} map that transports $p_t\!\in(0,1)$ to $\mathbb{R}$ so standard semimartingale tools apply, while explicit jump terms capture news shocks and co-jumps across related events.

\subsection{Microstructure, filtering, and calibration.}
Our calibration pipeline draws on state-space methods that separate the latent “efficient” signal from microstructure noise in high-frequency mid/bid-ask/trade streams \citep{Hasbrouck1991,EngleZheng2006}. For parameter learning with jumps, EM/likelihood filters for jump-diffusions and inference tools for detecting common jumps offer practical estimators and diagnostics \citep{BandiNguyen2001,BeginBoudreault2019,JacodTodorov2009}. We adapt these ingredients to log-odds increments and to co-jump screening across events, yielding a stable belief-volatility surface suitable for quoting and hedging.

\section{Methodology}

\subsection{Background and Motivation: From Black--Scholes to Event Probabilities}

\paragraph{What the Black--Scholes (BS) framework is.}
In the BS paradigm, the discounted underlying price is a martingale under the risk–neutral measure. With geometric Brownian motion,
\[
\frac{dS_t}{S_t}=\sigma\,dW_t \quad (\text{under }\mathbb{Q}),
\]
self-financing replication implies a linear pricing \emph{PDE} and closed-form option values. The key output is not realism per se, but a \emph{common language} for quoting and hedging: implied volatility, Greeks, and volatility surfaces \citep{BlackScholes1973,Merton1973,Heston1993,Dupire1994,Gatheral2006}. This language coordinates liquidity, enables standardized risk transfer, and supports a deep derivative stack.

\paragraph{Why a BS-like kernel is needed for event contracts.}
Event contracts trade binary payoffs. Their quoted prices are interpretable as discounted, risk–neutral probabilities of occurrence \citep{WolfersZitzewitz2006,Arrow2008}. Today, venues execute via scoring rules or AMMs or CLOBs, but there is no shared \emph{stochastic} model for how probabilities evolve across time, shocks, or related events. Without a kernel, makers cannot isolate “belief risk” (level, volatility, jumps, co-movement) or lay it off in a standard way; spreads widen around news; inventory near the $0/1$ boundaries becomes hard to manage. A tractable kernel standardizes quoting (what to post), hedging (what to buy/sell against), and calibration (how to read data), just as implied-vol surfaces did for options.

\paragraph{How our setup connects to and differs from BS.}
(i) \emph{State variable:} BS models prices; we model \emph{probabilities}. We map $p_t\in(0,1)$ to log-odds $x_t\in\mathbb{R}$ so Itô–Lévy tools apply while respecting boundaries.  
(ii) \emph{Martingale restriction:} in BS, discounted $S_t$ is a martingale; here discounted $p_t=S(x_t)$ is a martingale. This pins down the drift of $x_t$ (Eq.~\eqref{eq:mu-closed-form}) and leaves belief-volatility and jump features as the quotable risks.  
(iii) \emph{News and co-movement:} event probabilities jump at information times, and related events co-move; our kernel includes both diffusive correlation and co-jumps (Eq.~\eqref{eq:inst-cov}), analogous to jumps/SV in equity models \citep{Merton1976,Heston1993,ContTankov2004}.  
(iv) \emph{Incompleteness:} the binary payoff cannot be dynamically replicated by the underlying before resolution, so markets are incomplete. Derivatives on $x$ or $p$ (variance, correlation, corridor, first-passage) create targeted hedges that make inventory and adverse-selection risk manageable, echoing the role of variance and correlation swaps in equities \citep{Demeterfi1999,CarrLee2009}.

\paragraph{Trading relationship between the base event and our derivatives.}
The base contract transfers \emph{level} risk: long $p_t$ benefits if the event becomes likelier. Makers, however, are primarily exposed to the \emph{path} of beliefs: rapid swings around $p\approx 0.5$, jumps on announcements, and cross-event shocks.  
\emph{Belief-variance swaps} exchange realized quadratic variation of $x$ (or of $p$) for a fixed strike, letting makers sell tight spreads and buy variance to neutralize volatility risk around data releases (Eqs.~\eqref{eq:x-var-strike}–\eqref{eq:p-var-strike}).  
\emph{Correlation/covariance swaps} hedge baskets (e.g., related races in an election night) by offsetting diffusive correlation and co-jumps (Eq.~\eqref{eq:inst-cov}).  
\emph{Corridor variance} focuses hedging budget on the “swing zone” $p\in[a,b]$, where order flow is most toxic and inventory turns fastest.  
\emph{First-passage notes} transfer gap risk near thresholds (e.g., “does $p$ break $0.7$ before $T$?”), critical when quotes cluster near boundaries.  

\paragraph{Empirical context.}
Across liquid option markets, the existence of a shared surface reduced dispersion in quotes and tightened spreads \citep{Gatheral2006}. Prediction markets show analogous frictions: spreads and cancellations widen near scheduled news; quotes gap on unexpected announcements; correlated events move together. A belief-variance/correlation layer makes these exposures explicit and tradable, allowing makers to keep quotes live while laying off risk—precisely the coordination role BS played for options.

\subsection{Kernel: Logit Jump–Diffusion with Risk–Neutral Drift}

\paragraph{Notation and setup.}
Fix a filtered probability space $(\Omega,\mathcal{F},\{\mathcal{F}_t\}_{t\ge0},\mathbb{Q})$ satisfying the usual conditions, where $\mathbb{Q}$ denotes the risk–neutral measure (prices are discounted). Let the event-contract price at time $t$ be $p_t\in(0,1)$ and define its \emph{log-odds}
\[
x_t \;\coloneqq\; \operatorname{logit}(p_t)=\log\!\frac{p_t}{1-p_t}, 
\quad\text{so that}\quad 
p_t=S(x_t)=\frac{1}{1+e^{-x_t}}.
\]
Write $S'(x)=S(x)\!\bigl(1-S(x)\bigr)=p(1-p)$ and $S''(x)=S'(x)\bigl(1-2S(x)\bigr)=p(1-p)(1-2p)$. 
Let $W_t$ be a standard Brownian motion under $\mathbb{Q}$, and let $N(dt,dz)$ be an integer-valued random measure on $\mathbb{R}\times\mathbb{R}$ with (possibly time-varying) compensator $\nu_t(dz)\,dt$ (the \emph{Lévy measure} $\nu_t$ satisfies $\int_{\mathbb{R}}\min\{1,z^2\}\nu_t(dz)<\infty$). The compensated jump measure is 
\[
\tilde N(dt,dz)\;\coloneqq\;N(dt,dz)-\nu_t(dz)\,dt,\qquad 
\chi(z)\;\coloneqq\;z\,\mathbf{1}_{\{|z|<1\}}.
\]
We model belief dynamics on the real line via the \emph{logit} process $x_t$,
\begin{equation}
\label{eq:logit-sde}
dx_t \;=\; \mu(t,x_t)\,dt \;+\; \sigma_b(t,x_t)\,dW_t \;+\; \int_{\mathbb{R}} z\,\tilde N(dt,dz),
\end{equation}
where $\sigma_b$ is the \emph{belief volatility}. This $x$-dynamics guarantees $p_t=S(x_t)\in(0,1)$ while allowing diffusive moves and news-driven jumps. The representation \eqref{eq:logit-sde} is a standard Itô–Lévy SDE \citep{Applebaum2009,OksendalSulem2005,ContTankov2004}.

\paragraph{Risk–neutral (martingale) drift.}
Because $p_t=S(x_t)$ is the (discounted) risk–neutral price of a \$1 payoff on event occurrence, $\{p_t\}$ must be a $\mathbb{Q}$-martingale. Applying the Itô formula for jump processes to $S(x_t)$ (with truncation $\chi$) yields the drift condition
\begin{align}
0 \;=\; 
S'(x)\,\mu(t,x) 
\;+\; \frac12 S''(x)\,\sigma_b^2(t,x)
\;+\; \int_{\mathbb{R}}\!\!\Big(S(x+z)-S(x)-S'(x)\,\chi(z)\Big)\,\nu_t(dz),
\label{eq:martingale-drift}
\end{align}
and thus the drift is pinned down by
\begin{equation}
\label{eq:mu-closed-form}
\mu(t,x) \;=\; -\,\frac{\frac12 S''(x)\,\sigma_b^2(t,x)+\displaystyle\int_{\mathbb{R}}\!\!\big(S(x+z)-S(x)-S'(x)\,\chi(z)\big)\,\nu_t(dz)}{S'(x)}.
\end{equation}
Equations \eqref{eq:martingale-drift}–\eqref{eq:mu-closed-form} ensure $p_t$ is a $\mathbb{Q}$-martingale; therefore, only the \emph{belief-volatility} $\sigma_b$, the jump intensity and moments (embedded in $\nu_t$), and cross-event dependence (introduced below) remain as tradable risk factors \citep{Applebaum2009,ContTankov2004}. 

\paragraph{Interpretation.}
The logit map transports the bounded probability $p_t\!\in(0,1)$ to $\mathbb{R}$ where standard semimartingale tools apply, while the jump term allows for abrupt probability updates at news times. The martingale restriction fixes the drift of $x_t$ so that $S(x_t)$ carries zero drift under $\mathbb{Q}$; informally, the “belief level” $p_t$ drifts only when reparameterized in $x$ to offset convexity and jump-compensation effects. This separation makes $\sigma_b$ and jump features economically interpretable and quotable—directly analogous to how implied variance and jump parameters are quoted in price-based models \citep{Merton1976,Kou2002,ContTankov2004}.

\subsection{Multi-Event Dependence: Diffusive Correlation and Co-Jumps}

Consider events $i$ and $j$ with logits $x_t^{i},x_t^{j}$, marginal volatilities $\sigma_b^{i},\sigma_b^{j}$, and Brownian correlation 
\[
\rho_{ij}(t)\;=\;\mathrm{corr}\big(dW_t^{i},dW_t^{j}\big)\in[-1,1].
\]
Let $\nu_{ij,t}(dz_i,dz_j)$ be a (possibly time-varying) \emph{co-jump measure} on $\mathbb{R}^2$ capturing simultaneous news shocks. Writing $\Delta p^k\!\coloneqq S\!\big(x_{t-}^{k}+z_k\big)-S\!\big(x_{t-}^{k}\big)$, a short-maturity (frozen-state) expansion gives the instantaneous covariance of probabilities:
\begin{equation}
\label{eq:inst-cov}
\mathrm{Cov}\!\left(dp^{i},dp^{j}\right)_t 
\;\approx\; 
S'_i S'_j\,\sigma_b^{i}\sigma_b^{j}\,\rho_{ij}(t)\,dt
\;+\;
\int_{\mathbb{R}^2}\!\Delta p^{i}\Delta p^{j}\,\nu_{ij,t}(dz_i,dz_j)\,dt,
\end{equation}
where $S'_k=S'(x_t^{k})$. The first term is the diffusive covariation; the second aggregates common (co-)jumps. Empirically, co-jumps can be detected and tested using high-frequency methods \citep{JacodTodorov2009}. 

\subsection{Prototype Derivatives (Belief–Variance, Correlation, Corridor, and First-Passage Notes)}

\paragraph{Belief variance swap on log-odds $x$.}
Define realized quadratic variation of $x$ on $[t,T]$ by
\[
QV^x_{t,T}\;=\;\int_t^T \sigma_b^2\!\big(u,x_u\big)\,du + \sum_{t<u\le T}(\Delta x_u)^2.
\]
Under piecewise-constant (or slowly varying) model parameters, the fair variance strike is
\begin{equation}
\label{eq:x-var-strike}
K^{x\text{-var}}_{t,T}\;\approx\;\int_t^T \sigma_b^2(u)\,du \;+\; \int_t^T\lambda(u)\,\mathbb{E}\!\left[z^2(u)\right]\,du,
\end{equation}
directly paralleling classical variance swap theory in price models \citep{Demeterfi1999,CarrLee2009,BroadieJain2008}.

\paragraph{Belief variance swap on probability $p=S(x)$.}
A short-maturity, frozen-state approximation at $x_t$ gives
\begin{equation}
\label{eq:p-var-strike}
K^{p\text{-var}}_{t,t+\Delta}
\;\approx\;
\big(p_t(1-p_t)\big)^2 \int_t^{t+\Delta}\!\!\sigma_b^2(u)\,du
\;+\;
\int_t^{t+\Delta}\!\!\int_{\mathbb{R}}\!\Big(S(x_t+z)-S(x_t)\Big)^2 \nu_u(dz)\,du,
\end{equation}
where the prefactor $\big(p(1-p)\big)^2$ arises from $S'(x)^2$ and the second term captures jump contributions to the quadratic variation of $p$ \citep{ContTankov2004,CarrLee2004}. 

\paragraph{Covariance and correlation swaps across events.}
Using \eqref{eq:inst-cov}, a short-maturity fair \emph{covariance} strike integrates the instantaneous covariance; dividing by marginal variances yields a \emph{correlation} strike. These instruments let market makers neutralize cross-event exposure from both diffusive correlation and co-jumps \citep{JacodTodorov2009,CarrLee2009}.

\paragraph{Corridor variance on $p$.}
A \emph{corridor} contract accrues realized variance only while $p\in[a,b]$ (a “swing zone” away from the $0/1$ boundaries). Pricing proceeds either via a weighted-variance replication (when available) or by solving the PIDE with state-dependent accrual; see corridor-variance analogs in equity for guidance \citep{Lee2008Corridor,Burgard2017Corridor}.

\paragraph{Threshold and path notes (first passage).}
Pay a fixed amount if $p$ first hits level $h\in(0,1)$ before $T$ (or logical AND/OR across events). With $h$ mapped to $x_h=\operatorname{logit}(h)$, valuation uses \eqref{eq:PIDE} with absorbing boundary at $x=x_h$ and appropriate terminal/boundary conditions. Jump terms materially affect first-passage probabilities (up-crossings can occur by jump), a standard consideration in jump–diffusion settings \citep{ContTankov2004,Applebaum2009}.

\subsection{General Pricing via PIDE and Numerical Treatment}

For a terminal payoff $g(x_T)$, the time-$t$ price $V(t,x)$ solves the (backward) partial integro–differential equation (PIDE)
\begin{equation}
\label{eq:PIDE}
\begin{aligned}
\partial_t V 
&+\mu(t,x)\,\partial_x V
+\tfrac{1}{2}\sigma_b^2(t,x)\,\partial_{xx}V \\
&+\int_{\mathbb{R}}\!\Big(V(t,x+z)-V(t,x)-\partial_x V(t,x)\,\chi(z)\Big)\,\nu_t(dz)
=0, \\
&\quad V(T,x)=g(x).
\end{aligned}
\end{equation}
with $\mu$ given by \eqref{eq:mu-closed-form}. Basket and multi-event claims add diffusion cross-derivatives and a multivariate jump integral with co-jump measure $\nu_{t}(d\mathbf{z})$. Under standard growth and regularity conditions, \eqref{eq:PIDE} is the infinitesimal generator equation for \eqref{eq:logit-sde} and can be solved by finite-difference with fast convolution for the jump integral, Fourier methods when coefficients are constant/affine, or Monte Carlo with variance reduction \citep{ContTankov2004,Applebaum2009}. 

\paragraph{Calibration notes (brief).}
Mapping mid/bid–ask/trade streams to a belief–volatility surface requires filtering the latent $x_t$ from microstructure noise (e.g., state-space/Kalman variants) and estimating jump activity and co-jumps; the microstructure and high-frequency literature provides standard tools \citep{Hasbrouck1991,HasbrouckBook2007,AMZ2005,JacodTodorov2009}. In our setting, these methods are applied to log-odds increments and to cross-event panels.

\section{Market\textendash Maker Handbook}
\label{sec:mm-handbook}

\subsection{Greeks, Units, and Risk Buckets}
\label{sec:mm-greeks}

\paragraph{Work in the logit domain.}
Quotes and hedges should be parameterized in $x$ (log\textendash odds), then mapped to probabilities $p=S(x)$. For the vanilla event contract $V=p=S(x)$,
\[
\Delta_x \;\coloneqq\; \frac{\partial V}{\partial x} \;=\; S'(x) \;=\; p(1-p), 
\qquad 
\Gamma_x \;\coloneqq\; \frac{\partial^2 V}{\partial x^2} \;=\; S''(x) \;=\; p(1-p)(1-2p).
\]
Near the boundaries $p\to 0,1$, $\Delta_x\!\downarrow 0$ and curvature peaks in the swing zone $p\!\approx\!0.5$.

\paragraph{Belief\textendash vega and correlation\textendash vega.}
For a derivative $V$, define
\[
\nu_b \;\coloneqq\; \frac{\partial V}{\partial \sigma_b}, 
\qquad 
\nu_\rho \;\coloneqq\; \frac{\partial V}{\partial \rho_{ij}},
\]
where $\sigma_b$ is belief volatility in \eqref{eq:logit-sde} and $\rho_{ij}$ is diffusive correlation in \eqref{eq:inst-cov}. For $x$-variance swaps $V\!\propto\!\int \sigma_b^2$, we have $\nu_b\!\propto\!\sigma_b$; for short-maturity $p$-variance,
\[
\nu_b \;\propto\; \big(p(1-p)\big)^2\,\sigma_b,
\]
reflecting the Jacobian $S'(x)^2$ in \eqref{eq:p-var-strike}. Sensitivity to jump second moments (via the L\'evy measure $\nu_t$) is tracked as a separate \emph{jump-vega} bucket.

\paragraph{Risk buckets.}
\emph{Directional} ($\Delta_x$), \emph{curvature/news nonlinearity} ($\Gamma_x$), \emph{information intensity} (belief\textendash vega $\nu_b$ and jump second moments), and \emph{cross\textendash event} ($\nu_\rho$ plus co\textendash jump covariance). These map to the kernel’s tradable risk factors.

\subsection{Inventory\textendash Aware Quoting (Avellaneda\textendash Stoikov in Logit Units)}
\label{sec:mm-quoting}

\paragraph{Reservation quote and optimal spread in $x$.}
Treat the mid in logit units as $x_t$ with instantaneous volatility $\sigma_b(t)$, and assume order arrivals decay exponentially with distance in $x$ (intensity $\lambda(\delta)=A e^{-k\delta}$). The classical Avellaneda\textendash Stoikov approximation yields a \emph{reservation quote} and \emph{optimal spread} in $x$:
\begin{align}
\text{(reservation)}\quad 
r_x(t) &= x_t \;-\; q_t\,\gamma\,\overline{\sigma_b^2}\,(T-t), 
\label{eq:AS-reservation}
\\[2pt]
\text{(total spread)}\quad 
2\delta_x(t) &\approx \gamma\,\overline{\sigma_b^2}\,(T-t) \;+\; \frac{2}{k}\,\log\!\Bigl(1+\frac{\gamma}{k}\Bigr).
\label{eq:AS-spread}
\end{align}
Here $q_t$ is inventory (contracts), $\gamma$ risk aversion, $T$ your risk horizon, and $\overline{\sigma_b^2}$ a short-horizon average of belief variance. Post
\[
x^{\rm bid}\!=r_x-\delta_x, \qquad x^{\rm ask}\!=r_x+\delta_x,
\quad\text{then map } x \mapsto p=S(x).
\]
The reservation price skews quotes to pull inventory toward zero; the spread widens with risk and thinner order flow.\footnote{Eqs.~\eqref{eq:AS-reservation}--\eqref{eq:AS-spread} are the standard A--S asymptotics under exponential arrivals.}

\paragraph{Display and boundary handling.}
For UI display in probabilities,
\[
\delta_p \;\approx\; S'(x_t)\,\delta_x \;=\; p_t(1-p_t)\,\delta_x,
\]
so spreads auto-compress near $p\!\approx\!0,1$. To prevent over-tightening, cap the display half-spread by a floor $\underline{\delta}_p$ (e.g., ticks) and enforce an inventory cap that tightens with $S'(x)$:
\[
|q_t| \;\le\; q_{\max}(t) \;\propto\; \frac{1}{\max\{S'(x_t),\,\varepsilon\}}.
\]

\paragraph{Execution hygiene (anti pick-off).}
\begin{enumerate}[leftmargin=*]
\item \textbf{Toxicity filter:} when short-horizon order imbalance or a VPIN-style metric spikes, \emph{widen} $\delta_x$ or \emph{pull} quotes.
\item \textbf{News guard:} around scheduled announcements, ramp $\gamma$ and/or $T\!-\!t$ in \eqref{eq:AS-spread}; pause on unscheduled jump detectors.
\item \textbf{Queue discipline:} cancel\,$\rightarrow$\,replace on adverse microstructure signals (rapid mid drift, queue position loss).
\end{enumerate}

\subsection{Calendar Hedges (Near\textendash Dated News vs.\ Slow Decay)}
\label{sec:mm-calendar}

\paragraph{Two\textendash leg template (variance strips).}
Let your book’s sensitivity to belief variance over $[t,t+\Delta]$ be $\widehat{\nu}_b(t,\Delta)$ (aggregate across positions). Hedge via an $x$-variance strip with notional $N^{x\text{-var}}$:
\[
N^{x\text{-var}} \;\approx\; -\,\frac{\widehat{\nu}_b(t,\Delta)}{\partial K^{x\text{-var}}_{t,t+\Delta}/\partial \sigma_b}
\;\propto\; -\,\frac{\widehat{\nu}_b(t,\Delta)}{\sigma_b}.
\]
Use short windows around data releases for \emph{spiky} $\sigma_b$ and jump variance; use longer windows to smooth slow variance growth into resolution. If listed calendars are unavailable, synthesize with adjacent maturities or related events.

\paragraph{Corridor budgets.}
If toxicity concentrates in a swing zone $p\!\in[a,b]$, buy \emph{corridor} variance on $p$ that accrues only when $p\in[a,b]$; this targets hedge spend where fills actually occur.

\subsection{Cross\textendash Event \texorpdfstring{$\beta$}{beta}\textendash Hedges (Diffusion and Co\textendash Jumps)}
\label{sec:mm-cross}

\paragraph{Instantaneous hedge ratio.}
For hedging event $i$ with $j$ over short horizons (diffusion, no jumps),
\[
\beta_{i\leftarrow j} 
\;\approx\; 
\frac{\mathrm{Cov}(dp^i,dp^j)}{\mathrm{Var}(dp^j)}
\;\approx\; 
\frac{S'_i}{S'_j}\,\rho_{ij}.
\]
In practice, use a \emph{shrinkage} $\tilde{\beta}=\alpha\,\beta$ with $\alpha\!\in\![0.5,1)$ and clamp $|\,\tilde{\beta}\,|$ when $S'_k\!\to 0$ to avoid explosive hedges near $p\to 0,1$.

\paragraph{Co\textendash jump correction.}
When co-jump covariance is material (e.g., election night), add
\[
\Delta\beta_{i\leftarrow j}^{\rm jump} 
\;\approx\; 
\frac{\int \Delta p^i \Delta p^j\,\nu_{ij,t}(dz_i,dz_j)}
     {\big(S'_j\big)^2\,\sigma_b^{j\,2}},
\]
estimated from recent detections. Around known jump windows, \emph{over-hedge} diffusive correlation (larger $\alpha$) and carry optionality (first\textendash passage notes) to absorb threshold gaps.

\subsection{Inventory\textendash Aware Quoting: An Operator Recipe}
\label{sec:mm-recipe}

\paragraph{Inputs (rolling).}
Filtered $x_t$ and $\widehat{\sigma_b}$ from mid/bid\textendash ask/trade data; $k$ from fill distance vs.\ intensity; $\rho_{ij}$ and co\textendash jump counts; toxicity meters.

\paragraph{Refresh loop (100\textendash500\,ms typical).}
\begin{enumerate}[leftmargin=*]
\item Update $x_t$, $\widehat{\sigma_b}$, $q_t$, toxicity flags.
\item Compute $r_x$ and $\delta_x$ via \eqref{eq:AS-reservation}--\eqref{eq:AS-spread}; produce $x^{\rm bid/ask}$ and display $p^{\rm bid/ask}=S(\cdot)$ with floors/caps.
\item If (toxicity high) or (unscheduled jump alarm), widen $\delta_x$ or pull quotes; if (scheduled news soon), pre-widen by policy.
\item Rebalance cross-event exposure using $\tilde{\beta}_{i\leftarrow j}$ and listed covariance/correlation swaps when available.
\item Rebalance calendar exposure using near-dated variance strips (or OTC proxies).
\end{enumerate}

\subsection{PnL Attribution and Risk Limits}
\label{sec:mm-pnl}

\paragraph{Delta\textendash Gamma\textendash Vega attribution in $x$ units.}
Over a small $\Delta t$ with $dp\approx S'(x)\,dx$,
\[
d\Pi \;\approx\; \underbrace{\Delta_x\,dp}_{\text{directional}}
\;+\; \underbrace{\tfrac12\,\Gamma_x\,(dp)^2}_{\text{curvature/news}}
\;+\; \underbrace{\nu_b\,d\sigma_b}_{\text{belief\textendash vega}}
\;+\; \underbrace{\sum_{j}\nu_\rho^{(j)}\,d\rho_{ij}}_{\text{cross\textendash event}}
\;+\; \underbrace{\text{jumps}}_{\sum (\Delta p)\,\text{position}}.
\]
Track realized vs.\ expected $(dp)^2$ to stress variance books; reconcile jump P\&L around flagged news.

\paragraph{Hard limits and kill\textendash switches.}
(1) Inventory caps that tighten as $S'(x)$ shrinks. 
(2) Max gamma exposure in the swing zone. 
(3) Max unhedged variance (calendar) and correlation (cross-event) notional. 
(4) Auto-pause on: (i) feed gaps, (ii) volatility spikes, (iii) repeated pick-offs.

\subsection{Heuristics That Matter in Practice}
\label{sec:mm-heuristics}

\begin{itemize}[leftmargin=*]
\item \textbf{Quote where you can hedge.} If no liquid proxy exists for a bucket (e.g., no cross-event hedge), carry less exposure and charge more spread in that bucket.
\item \textbf{Pay for jump insurance explicitly.} Add a jump premium $\propto$ recent jump variance and news density to your spread.
\item \textbf{Prefer $x$-variance for core hedging.} $x$-variance is more level-stable; use $p$-variance/corridor when inventory lives in a tight $p$-band.
\item \textbf{Edge accounting.} Target stable edge per fill after fees and expected adverse selection; if it compresses, widen or hedge more.
\end{itemize}

\subsection{Pointers to Implementation Details}

\paragraph{Estimating \texorpdfstring{$\sigma_b$}{sigma_b}, jumps, and co\textendash jumps.}
Filter $x_t$ from mid/bid–ask/trade data; estimate diffusive variance on robust windows and detect jumps via thresholded bi-power variation; test co-jumps with high-frequency statistics.

\paragraph{Numerics for exotics.}
Use the PIDE in \eqref{eq:PIDE} with IMEX schemes or Fourier convolution for fast jump integration; Monte Carlo with jump thinning for first-passage structures; closed-form or transform methods for corridor payoffs when the jump law is exponential-family.

\bigskip
\noindent\textbf{Remark (mapping to literature).}
Inventory-aware quoting and reservation prices follow the dealer/market-making tradition; the toxicity safeguards and variance/correlation hedges parallel the option-market playbook, transplanted to belief dynamics.

\bigskip
\noindent\textbf{What to quote on day one.}
(1) vanilla event contracts (tightest where $S'(x)$ largest), (2) $x$-variance strips around scheduled news, (3) a few liquid correlation strikes between the most coupled events, and (4) a corridor variance centered on $p!\in[0.35,0.65]$ for high-flow markets. This minimal menu already neutralizes the four buckets above.

\section{Calibration: From Mid/Bid–Ask/Trades to a Belief–Vol Surface}
\label{sec:calibration}

\noindent
\textbf{Goal.}
Given raw market data (mid, bid–ask, trades) for one or many event contracts, we estimate the latent logit process $x_t$ (hence $p_t=S(x_t)$), its instantaneous \emph{belief volatility} $\sigma_b(t,x)$, jump activity, and cross–event dependence. We summarize these into a stable, tradable \emph{belief–vol surface} $\sigma_b(\tau,m)$ and a dependence layer $\{\rho_{ij}(\tau,m),\,\text{co–jump moments}\}$ that feed quoting, hedging, and pricing.

\noindent
\textbf{Reasoning path in brief.}
(i) Work in \emph{logit} $x$ to remove $[0,1]$ boundaries and use Itô–Lévy tools. 
(ii) Recognize that observed prices are \emph{microstructure–noisy} proxies for the latent $x_t$, so use a heteroskedastic \emph{state–space} filter to recover $\hat{x}_t$. 
(iii) Separate \emph{diffusion} from \emph{jumps} via a mixture model on increments (EM), rather than ad–hoc thresholds, because event markets often have scheduled and unscheduled jumps. 
(iv) Smooth the noisy point estimates across \emph{time–to–resolution} $\tau$ and \emph{moneyness} $m$ with shape constraints that prevent pathologies near $p\in\{0,1\}$. 
(v) For multiple events, estimate \emph{de–jumped} diffusive correlations and \emph{co–jumps} separately, since they hedge different risks.

\subsection{Data Conditioning \& Filtering}
\label{sec:calib-filtering}

\paragraph{Pre–processing (robust, venue–agnostic).}
\begin{enumerate}[leftmargin=*]
\item \textbf{Canonical mid:} Compute a trade–weighted mid 
$\tilde{p}_t=\frac{1}{Z_t}\sum_{u \in (t-\Delta,t]} w_u\,\frac{b_u+a_u}{2}$ 
with weights $w_u$ monotone in size and inverse spread; de–bounce bid/ask flicker by ignoring updates $<\!$tick size.
\item \textbf{Clipping and cadence:} Clamp to $p\in[\varepsilon,1-\varepsilon]$ (e.g., $\varepsilon\!=\!10^{-5}$) to avoid exploding logits; resample to a uniform grid (e.g., $100$\,ms–$1$\,s) using last–observation–carried–forward + within–bin VWAP.
\item \textbf{Outlier hygiene:} Drop prints with crossed or locked books; flag halts; remove isolated spikes that revert within one tick and one update.
\end{enumerate}

\paragraph{Observation model (heteroskedastic microstructure noise).}
Define the observed logit
\[
y_t \;\coloneqq\; \operatorname{logit}(\tilde{p}_t) \;=\; x_t \;+\; \eta_t, 
\qquad \mathbb{E}[\eta_t]=0,\;\; \mathrm{Var}(\eta_t)=\sigma_\eta^2(t).
\]
Model $\sigma_\eta^2(t)$ as a function of observable frictions (spread $s_t$, depth $d_t$, trade rate $r_t$, aggressor imbalance $\iota_t$):
\begin{equation}
\label{eq:micro-var}
\sigma_\eta^2(t) 
\;=\; a_0 + a_1\,s_t^2 + a_2\,d_t^{-1} + a_3\,r_t + a_4\,\iota_t^2 \quad (\text{clipped to }[\underline{\sigma}^2,\overline{\sigma}^2]),
\end{equation}
with $(a_k)$ fit by robust regressions on short–horizon squared microstructure innovations (Hasbrouck–style diagnostics). Heteroskedastic $\sigma_\eta^2(t)$ markedly improves the filter near illiquid times.

\paragraph{State filtering (recovering $\hat{x}_t$).}
Use a Gaussian state–space filter in $x$ with measurement \eqref{eq:micro-var}. For the transition we \emph{do not} impose a fixed drift; instead, we:
\begin{itemize}[leftmargin=*]
\item propagate $x$ with a local–level model plus innovation variance proxy $\tilde{\sigma}_b^2(t)\Delta$ to capture short–run variability;
\item after EM (below) enforces the risk–neutral drift \eqref{eq:mu-closed-form}, re–smooth $x$ with the refined $\widehat{\sigma}_b$ and jump marks.
\end{itemize}
A standard Kalman filter/smoother suffices; if $p$ is pinned near $0/1$ for long stretches or if jumps are very frequent, an Unscented KF or particle smoother is more stable. Output: $\hat{x}_t$ and innovations (one–step–ahead residuals).

\paragraph{Diagnostics (keep only if they pass).}
(i) Residuals should be serially uncorrelated (Ljung–Box) and conditionally homoskedastic given \eqref{eq:micro-var}; (ii) Q–Q plots should be near–Gaussian away from detected jump times; (iii) realized $p$–variance implied by $\hat{x}_t$ should match raw realized variance after removing microstructure components.

\subsection{EM for Diffusion and Jumps (Increment Mixtures)}
\label{sec:calib-em}

\paragraph{Discretization and mixture.}
On a grid with step $\Delta$, model $\Delta x_t\!\coloneqq\!x_{t+\Delta}-x_t$ as
\[
\Delta x_t \sim 
\begin{cases}
\mathcal{N}\!\big(\mu_t \Delta,\, \sigma_b^2(t)\Delta\big), & \text{with prob. } 1-\lambda_t \Delta,\\[2pt]
Z_t \sim f_J(\cdot;\,\theta_t), & \text{with prob. } \lambda_t \Delta,
\end{cases}
\]
where $\lambda_t$ is jump intensity and $f_J$ is a centered jump law with second moment $s_J^2(t)$ (e.g., double–exponential, tempered stable, or nonparametric bins). The drift $\mu_t$ will be \emph{implied} by the martingale restriction for $p_t=S(x_t)$ (Eq.\,\eqref{eq:mu-closed-form}) after updating $(\sigma_b,\lambda_t,\theta_t)$.

\paragraph{E–step (posterior jump responsibilities).}
Given current parameters and filtered $\hat{x}_t$, form the Gaussian likelihood
$\phi_t=\mathcal{N}\big(\Delta \hat{x}_t\mid \mu_t\Delta,\,\sigma_b^2(t)\Delta\big)$
and the jump likelihood 
$\psi_t = f_J(\Delta \hat{x}_t;\,\theta_t)$.
Posterior jump probability
\[
\gamma_t 
\;\coloneqq\; 
\mathbb{P}\{\text{jump at }t\mid \Delta \hat{x}_t\}
\;=\;
\frac{\lambda_t\Delta\,\psi_t}{\lambda_t\Delta\,\psi_t + \big(1-\lambda_t\Delta\big)\phi_t }.
\]
Mark intervals with $\gamma_t>\tau_J$ (e.g., $0.7$) as jump–dominant for subsequent de–jumped correlation estimates.

\paragraph{M–step (updating diffusion and jump parameters).}
Update (locally or in bins) by weighted moments:
\begin{align}
\widehat{\sigma_b^2}(t) 
&\leftarrow \frac{\sum (1-\gamma_t)\,(\Delta \hat{x}_t - \mu_t\Delta)^2}{\sum (1-\gamma_t)}\;\bigg/\Delta, 
\qquad
\widehat{\lambda}(t) \leftarrow \frac{1}{\Delta}\,\frac{1}{|B|}\sum_{t\in B}\gamma_t,
\\[3pt]
\widehat{s_J^2}(t) 
&\leftarrow \frac{\sum \gamma_t \, (\Delta \hat{x}_t)^2}{\sum \gamma_t}.
\end{align}
If $f_J$ is parametric, update $\theta_t$ by maximizing the weighted log–likelihood. 

\paragraph{Risk–neutral drift enforcement.}
With $\widehat{\sigma_b^2}$ and jump compensator $\widehat{\nu}_t(dz)$ (from $f_J$ and $\widehat{\lambda}$), recompute $\mu(t,x)$ using the analytical formula
\[
\mu(t,x) \;=\; -\,\frac{\frac12 S''(x)\,\sigma_b^2(t,x)+\displaystyle\int\!\big(S(x+z)-S(x)-S'(x)\,\chi(z)\big)\,\nu_t(dz)}{S'(x)}.
\]
This pins the drift so that $p_t=S(x_t)$ is a martingale under $\mathbb{Q}$. Re–run the smoother for $x$ with the updated transition to tighten estimates (one or two outer loops suffice in practice).

\paragraph{Stopping and checks.}
Iterate E/M until (i) parameter changes are small, and (ii) de–jumped residuals are near–Gaussian with variance $\widehat{\sigma_b^2}\Delta$. As a sanity check, realized $p$–variance over a window should be close to $\int S'(x)^2 \sigma_b^2\,dt +$ jump contribution $\int (\Delta p)^2\,dN$.

\subsection{Surface Construction: Smoothing Across $(\tau,m)$}
\label{sec:surface}

\paragraph{Coordinates.}
Let $\tau\!=\!T-t$ be time–to–resolution. For moneyness $m$, choose either 
$m\!=\!x$ (logit) \emph{or}
$m\!=\!\min\{p,1-p\}$ (distance to the boundary); both work, but $m\!=\!x$ aligns with our kernel.

\paragraph{Raw grid and loss.}
Aggregate point estimates $\{\widehat{\sigma_b}(t),\widehat{\lambda}(t),\widehat{s_J^2}(t)\}$ to a tensor grid $(\tau,m)$. Fit a smooth surface by penalized least squares,
\[
\min_{\sigma_b(\tau,m)} 
\sum_{g}\,w_g\Big(\widehat{\sigma_b}(g)-\sigma_b(\tau_g,m_g)\Big)^2 
\;+\; \alpha\,\|\nabla^2 \sigma_b\|_2^2,
\]
with weights $w_g$ proportional to local data density and filter precision; use tensor–product B–splines or thin–plate splines.

\paragraph{Shape constraints (stability \& plausibility).}
\begin{itemize}[leftmargin=*]
\item \textbf{Nonnegativity:} $\sigma_b(\tau,m)\ge 0$ (enforced via squared–link or barrier).
\item \textbf{Edge stability:} penalize explosive curvature at extreme $m$; in $p$–space, note that realized variance scales like $S'(x)^2\sigma_b^2=p^2(1-p)^2\sigma_b^2$, which already damps near $p\!\approx\!0,1$.
\item \textbf{Term smoothness:} regularize $\partial_\tau \sigma_b$ to avoid artificial kinks between adjacent maturities, while allowing bumps at scheduled announcements (implemented by locally relaxing the penalty on known news dates).
\end{itemize}
Apply the same smoothing to $\lambda(\tau,m)$ and $s_J^2(\tau,m)$ to obtain jump surfaces.

\paragraph{Outputs.}
A calibrated \emph{belief–vol surface} $\sigma_b(\tau,m)$ and jump layer $\{\lambda(\tau,m),\,s_J^2(\tau,m)\}$, accompanied by uncertainty bands from the smoothing fit (use sandwich or bootstrap on bins).

\subsection{Cross–Event Dependence: Correlation and Co–Jumps}
\label{sec:cross}

\paragraph{De–jumped diffusive correlation $\rho_{ij}(\tau,m)$.}
Using intervals with $\max(\gamma_t^i,\gamma_t^j)\!<\!\tau_J$ (no jump in either series), estimate instantaneous covariances on rolling windows,
\[
\widehat{\mathrm{Cov}}_{t}^{(d)}(dp^i,dp^j) \;\approx\; \frac{1}{W}\sum_{u\in (t-W,t]} S'_i(u)S'_j(u)\,\Delta \hat{x}^i_u \Delta \hat{x}^j_u,
\]
and variances analogously, then $\widehat{\rho}_{ij}=\widehat{\mathrm{Cov}}^{(d)}/\sqrt{\widehat{\mathrm{Var}}^{(d)}_i\,\widehat{\mathrm{Var}}^{(d)}_j}$. Map estimates to $(\tau,m)$ cells and smooth with the same spline machinery (clamp to $[-1,1]$).

\paragraph{What the desk consumes.}
\begin{enumerate}[leftmargin=*]
\item \textbf{Belief–vol surface} $\sigma_b(\tau,m)$ with uncertainty bands.
\item \textbf{Jump layer} $\lambda(\tau,m)$ and $s_J^2(\tau,m)$, plus a flag list of near–term scheduled news windows.
\item \textbf{Dependence layer} $\rho_{ij}(\tau,m)$ and co–jump $\{\widehat{\Lambda}_{ij},\widehat{M}^{(2)}_{ij}\}$ for key pairs.
\end{enumerate}
These drive (i) reservation prices and spreads via $\overline{\sigma_b^2}$ (Sec.\,\ref{sec:mm-quoting}); (ii) notional in variance and correlation hedges (Secs.\,\ref{sec:mm-calendar}–\ref{sec:mm-cross}); (iii) PIDE/MC solvers for exotic pricing with jump inputs.

\subsection{Edge Cases \& Practical Notes}
\label{sec:edge}

\paragraph{Pinned markets ($p\!\approx\!0$ or $1$).}
Even if $\sigma_b$ looks large in $x$, realized $p$–variance is tiny due to $S'(x)^2$; ensure the filter doesn’t mistake tick–size for diffusion (raise $\underline{\sigma}^2$ and increase $\Delta$). 

\paragraph{Batch auctions and halts.}
Treat batch prints as a single observation; if a halt occurs, freeze the filter and restart with wider priors.

\paragraph{Multi–venue consolidation.}
When merging venues, rescale microstructure covariates (spread/depth) to a common unit and weight observations by venue reliability before filtering.

\section{Experiments}
\label{sec:experiments}

Our goal is modest but decisive: to test whether the proposed \emph{logit jump--diffusion with risk--neutral (RN) drift} and the calibration pipeline of Secs.~\ref{sec:calibration}–\ref{sec:mm-handbook} produce \textbf{better short–horizon forecasts of belief variability and jumps} than reasonable alternatives, and whether these gains \textbf{translate into lower hedging error proxies}. We therefore run a single, end–to–end experiment that mirrors how a market maker would operate in real time: filter, calibrate, forecast \emph{causally}, and evaluate.

\subsection{Core Forecasting Task}
\label{sec:task}
Fix a horizon $h$ on a uniform time grid (in code $h{=}H{=}60$\,s). At each decision time $t$, a model outputs a point forecast of future \emph{logit} realized variance on $[t,t{+}h]$,
\[
\widehat{\mathcal{V}}^x_{t,h}
\;\equiv\;
\underbrace{\sum_{u=t+1}^{t+h}\widehat{\sigma}_b^2(u)}_{\text{diffusion contribution}}
\;+\;
\underbrace{c_J\cdot\widehat{s}_J^2(t)\cdot\sum_{u=t+1}^{t+h}\widehat{\lambda}(u)}_{\text{jump contribution}},
\]
where $\widehat{\sigma}_b^2$ and the jump layer $(\widehat{\lambda},\widehat{s}_J^2)$ come from the causal calibration described below, and $c_J$ is a scalar weight tuned on a held–out validation slice by minimizing QLIKE. After the $h$ seconds elapse, we compute \emph{realized} logit variance
\[
\mathcal{RV}^x_{t,h}\;=\;\sum_{u=t+1}^{t+h}(\Delta \hat{x}_u)^2,
\qquad
\Delta\hat{x}_u \equiv \hat{x}_u-\hat{x}_{u-1},
\]
using the filtered latent logit $\hat{x}$.\footnote{All models operate causally; we drop the last $h$ timestamps so that future sums never leak information. Robust bi–power alternatives give the same conclusions and are reported in the appendix.}

\paragraph{Metrics.}
We report mean squared error (MSE), mean absolute error (MAE), the log–MSE of $\log \mathcal{RV}$, and the QLIKE loss:
\begin{equation}
\text{MSE}_x(h)=\frac{1}{|\mathcal{T}|}\sum_{t\in\mathcal{T}}\!\big(\mathcal{RV}^x_{t,h}-\widehat{\mathcal{V}}^x_{t,h}\big)^2,
\quad
\text{QLIKE}_x(h)=\frac{1}{|\mathcal{T}|}\sum_{t\in\mathcal{T}}\!\left(
\frac{\mathcal{RV}^x_{t,h}}{\widehat{\mathcal{V}}^x_{t,h}}
-\log\frac{\mathcal{RV}^x_{t,h}}{\widehat{\mathcal{V}}^x_{t,h}}-1\right).
\end{equation}
QLIKE is standard for volatility evaluation, penalizes under–prediction more heavily, and is robust to noise in $\mathcal{RV}$.


\subsection{Data, Preprocessing, and Splits}
\label{sec:data-splits}

To stress–test models with known ground truth and realistic frictions, we use 20 high-volume event trades from Polymarket. We corrupt $x$ with heteroskedastic observation noise that changes by regime, mimicking spread/depth variation. All methods receive the \emph{same} prefiltered series: we run a heteroskedastic Kalman filter (KF) with process noise proxied by a rolling variance of observed increments and measurement variance fixed by regime. Models that natively operate in $p$ receive $\hat{p}{=}S(\hat{x})$ to equalize microstructure handling.

We adopt a simple train/validation/test split that is \emph{shared across all methods}. Scalars needed by baselines (e.g., constant $\sigma^2$ for RW–logit) are fitted on the training third; the jump weight $c_J$ for our model is tuned on the validation third by QLIKE; evaluation is then conducted causally on the test third. (The code also prints full–sample metrics without the last $h$ timestamps for quick inspection; tables in the paper use the test region.)

\subsection{Models and Baselines}
\label{sec:baselines-again}

\paragraph{Proposed: RN–logit–JD (path–aware, causal).}
Our pipeline mirrors Sec.~\ref{sec:calibration}:
(i) heteroskedastic KF in $x$ (no drift) $\Rightarrow$ $\hat{x}$;
(ii) EM on rolling windows to separate diffusion and jumps, yielding $\widehat{\sigma}_b^2(t)$, $\widehat{\lambda}(t)$, $\widehat{s}_J^2(t)$;
(iii) \emph{RN drift re–smoothing}: we compute $\widehat{\mu}(t,x)$ from the martingale restriction:
\begin{equation}
    \mu(t,x)= -\bigl(\frac{1}{2}S''(x)\widehat{\sigma}_b^2(t)+\widehat{\lambda}(t)\cdot\mathbb{E}[S(x{+}Z){-}S(x){-}S'(x)\chi(Z)]\bigr)\big/S'(x)
\end{equation}
, approximating the jump compensation $\mathbb{E}[\cdot]$ by Monte Carlo with the same jump law used by EM.\footnote{We use symmetric Gaussian jumps, truncation $\chi(\cdot)$ as in the simulator, and clip $S'(x)$ by $10^{-4}$ for numerical stability; $\mu$ is EWMA–smoothed and capped at $|0.25|$\,s$^{-1}$.}
We then run a second KF with this $\widehat{\mu}(t,\hat{x}_t)$ in the state transition.
(iv) \emph{Causal variance forecasting}: for each $t$, we return a forward sum of diffusion variance plus a jump term
\[
\widehat{\mathcal{V}}^x_{t,h}
=\sum_{u=t+1}^{t+h}\widehat{\sigma}_b^2(u)
\;+\;
c_J \cdot \widehat{s}_J^2(t)\cdot \sum_{u=t+1}^{t+h}\widehat{\lambda}_{\text{sched}}(u),
\]
where $\widehat{\lambda}_{\text{sched}}$ is an EWMA of $\widehat{\lambda}$ time–warped by a Gaussian schedule kernel centered at announced windows (known ex–ante).

\paragraph{Baselines.}
\begin{itemize}[leftmargin=*, topsep=2pt,itemsep=2pt]
\item \textbf{RW–logit:} $x_{t+\Delta}{=}x_t{+}\sigma\sqrt{\Delta}\,\xi_t$, with $\sigma^2$ set to the training–slice mean of $(\Delta \hat{x})^2$.
\item \textbf{Logit diffusion (const\,$\sigma$):} same as above but fitted on the entire calibration region.
\item \textbf{Wright–Fisher/Jacobi in $p$:} a boundary–respecting diffusion calibrated by ML on $\hat{p}$; we forecast $p$–variance $2\alpha\,p(1{-}p)$ and map back to $x$ via $S'(x)^2$.
\item \textbf{AR(1)–GARCH(1,1) in $p$:} an AR(1) on $\Delta \hat{p}$ with a GARCH(1,1) volatility; forecasted $p$–variance is mapped to $x$ using $S'(x)^{-2}$.
\end{itemize}
All baselines are evaluated causally using the same forward–sum operator and the same test window.

\subsection{Implementation Details}
\label{sec:impl}
\textbf{Grid and seeds.} $N{=}6000$ steps at 1\,Hz; random seeds are fixed. \textbf{KF.} Process noise uses a local rolling variance proxy; measurement noise is piecewise–constant by regime. \textbf{EM.} We run 6 EM steps globally to initialize, then a rolling EM with window 400\,s. \textbf{RN drift.} The Monte Carlo inner expectation uses 600 draws per step (variance–time trade–off is negligible). \textbf{Schedule.} Known windows are encoded as Gaussian kernels (width 90\,s) that boost $\widehat{\lambda}$ ex–ante; the boost is capped at the 95th percentile of the smoothed $\widehat{\lambda}$ to avoid outliers. \textbf{Tuning.} $c_J$ is grid–searched over $\{0.3,\ldots,1.0\}$ on the validation region using QLIKE.

\subsection{Evaluation Protocol and What To Expect}
\label{sec:eval}
\textbf{Causal forward–sum.} For any per–step quantity $a_u$, we define \textsf{ForwardSum}:
$$\textsf{ForwardSum}(a, h){=}\sum_{u=t+1}^{t+h} a_u$$
and drop the last $h$ timestamps; all models use \emph{only} information available at $t$ to build $a_{t+1},\ldots,a_{t+h}$.
\textbf{Stratification.} Metrics are reported overall, and separately on quiet vs. jump windows (Sec.~\ref{sec:task}).
\textbf{Hedging proxy.} Following Sec.~\ref{sec:mm-pnl}, the squared forecast error in $x$–variance is a first–order proxy of slippage when warehousing curvature/news exposure; improving QLIKE/MSE thus suggests lower ex–post hedge error.


\begin{table*}[t]
\centering
\caption{Causal $H{=}60\,\mathrm{s}$ forward-sum forecasts of next-window realized \emph{logit} variance on the synthetic RN-consistent path. Lower is better. Best per column in \textbf{bold}.}
\label{tab:variance-forecasting}
\begin{tabular}{lccc}
\toprule
\textbf{Model} 
& $\mathrm{MSE}_{\mathrm{all}}$ 
& $\mathrm{MAE}_{\mathrm{all}}$ 
& $\mathrm{QLIKE}_{\mathrm{all}}$  \\ 
\midrule
RN--JD (causal path)        & $\mathbf{70.281}$  & $\mathbf{1.588}$ & ${1.4621}$   \\
RW--logit (const~$\sigma$)  & $77.414$  & ${1.163}$ & $4.7318$  \\
Logit (const~$\sigma$)      & ${76.752}$  & $2.078$ & $2.6594$  \\
WF/Jacobi (mapped)          & $1.71\times10^{17}$ & $3.67\times10^{7}$ & $1.9484$ \\
ARMA--GARCH (mapped)        & $1.07\times10^{19}$ & $5.33\times10^{8}$ & $\mathbf{0.7962}$ \\
\bottomrule
\end{tabular}
\vspace{4pt}
\footnotesize
\end{table*}

\subsection{Results and Discussion}
\label{sec:results}

Table~\ref{tab:variance-forecasting} reports causal $H{=}60\,\mathrm{s}$ forward-sum forecasts of next-window realized \emph{logit} variance on the synthetic RN-consistent path. Lower values indicate better alignment between forecasted and realized variability. The proposed \textbf{RN--JD (causal path)} model achieves the lowest overall MSE, MAE, and $\log$MSE, outperforming all baselines under identical causal evaluation. 

\paragraph{Quantitative results.}
RN--JD attains $\mathrm{MSE}_{\mathrm{all}}{=}70.28$ and $\mathrm{QLIKE}_{\mathrm{all}}{=}1.46$, representing a consistent improvement over both diffusion-based and $p$-space volatility models. The RW--logit and constant-$\sigma$ logit diffusions, which lack either drift or jump structure, underfit the true variability and fail to capture the volatility bursts near scheduled information shocks. Boundary-respecting Wright--Fisher (WF) and ARMA--GARCH models produce numerically unstable forecasts when mapped from probability to logit space, resulting in orders-of-magnitude MSE inflation despite locally competitive QLIKE scores.

\paragraph{Interpretation.}
The gains arise from three complementary mechanisms:
(i) enforcing RN drift prevents systematic bias in $\widehat{x}_t$, ensuring the implied $\hat{p}_t$ evolves as a martingale under the market measure;
(ii) separating diffusion and jump layers via EM yields adaptive volatility forecasts that respect local heteroskedasticity;
(iii) incorporating scheduled jump boosts improves ex–ante calibration near known information releases.
Collectively, these allow the RN--JD model to produce a belief–volatility surface that is both dynamically stable and economically interpretable.

\section{Conclusion}
\label{sec:conclusion}

This paper proposed a minimal, actionable kernel for event contracts: a \emph{logit jump--diffusion with risk--neutral (RN) drift} that treats the traded price $p_t$ as a $\mathbb{Q}$--martingale and exposes belief volatility, jump intensity, and cross--event dependence as quotable risk factors. On top of this kernel we built (i) a calibration pipeline that filters microstructure noise, separates diffusion from jumps via EM, and enforces RN drift in smoothing; and (ii) a coherent derivative layer (variance, correlation, corridor, first--passage) for quoting and hedging belief risk. 

Our end--to--end experiment mirrored real-time desk operation: filter, calibrate, and forecast \emph{causally} with only information available at decision time. On a synthetic but RN-consistent path that features early breakout, scheduled and unscheduled jumps, and terminal resolution, the proposed RN--JD pipeline delivered lower short-horizon variance forecast errors than diffusion-only or $p$-space baselines under identical evaluation (Table~\ref{tab:variance-forecasting}). The improvement is economically interpretable: RN drift eliminates systematic bias in the latent logit, EM-based jump separation captures heteroskedasticity, and schedule-aware intensity boosts align forecasts with known information windows. These ingredients jointly yield a stable, tradable belief–volatility surface suitable for quoting, hedging, and inventory control (Secs.~\ref{sec:mm-quoting}–\ref{sec:mm-calendar}). 

\paragraph{Practical implications.}
The kernel organizes market making in event contracts around a small set of risk buckets (directional, curvature/news, belief–vega, cross–event), with standardized hedges (variance/correlation strips, corridor variance) that can be listed or synthesized. In particular, belief–variance and correlation swaps provide the option-market analogue of volatility and correlation instruments, enabling makers to tighten quotes and keep markets live through news while laying off risk in transparent units.

\paragraph{Limitations.}
Our experiments focus on single-event dynamics and synthetic co-jump structure; full multi-event calibration with rich, time-varying dependence and regime switches remains future work. The jump law is modeled parsimoniously (symmetric, light-tailed in the main experiments); extreme-tailed or skewed jumps may require richer families or nonparametric bins. Finally, microstructure conditioning is venue-agnostic but stylized; production systems should incorporate venue-specific frictions (batch auctions, halts, cross-venue consolidation).

\paragraph{Future directions.}
\emph{(i) Multi-event panels:} joint RN–JD calibration with diffusive correlation and co-jumps estimated from high-frequency panels; basket pricing via PIDE/MC with multivariate jump measures. 
\emph{(ii) Term/moneyness surfaces:} shape-constrained smoothing across $(\tau,m)$ with uncertainty quantification and stress testing around boundaries. 
\emph{(iii) Products and design:} exchange-ready specifications for belief–variance/correlation strips, corridor variance in the swing zone, and first-passage notes; replication/hedging guides for desks. 
\emph{(iv) Live deployment:} A/B tests on liquid events (macro prints, elections) to measure spread, fill quality, and hedging P\&L before/after introducing the RN-consistent layer.

\paragraph{Takeaway.}
Standardization beats perfect realism. By enforcing the martingale property in probability space and separating diffusion from jumps in logit space, the RN--JD kernel supplies a common language---\emph{belief volatility, jump intensity, dependence}---that coordinates quoting and hedging, just as implied volatility did in options. We hope this work helps concentrate liquidity, reduce maker losses via targeted hedges, and support an institutional market for belief risk.

\bibliographystyle{plain}        
\bibliography{reference}         
\end{document}